\documentclass{aa}  

\usepackage{amsmath}
\usepackage{tablefootnote}
\usepackage{bm}
\usepackage{graphicx}
\usepackage{txfonts}
\usepackage[colorlinks,linkcolor=blue,citecolor=blue,linktocpage=true,breaklinks, 
plainpages=false,urlcolor=blue]{hyperref}

\begin{document}

   \title{Using the infrared iron lines to probe solar subsurface convection}

   \author{I. Mili\'{c}
          \inst{1,2,3,4}
          \and
          H. N. Smitha
          \inst{1} 
          \and
          A. Lagg
          \inst{1}
          }

   \institute{Max Planck Institute for Solar System Research, Justus-von-Liebig Weg 3, 37077 G\"{o}ttingen, Germany\\
   \and
   Department of Physics, University of Colorado, Boulder CO 80309, USA\\
   \and
   Laboratory for Atmospheric and Space Physics, University of Colorado, Boulder CO 80303, USA\\
   \and
   National Solar Observatory, Boulder CO 80303, USA\\
   \email{ivan.milic@colorado.edu}
   }

   \date{Received ; accepted }

 
  \abstract
   {Studying the properties of the solar convection using high-resolution spectropolarimetry began in the early 90's with the focus on observations in the visible wavelength regions. Its extension to the infrared (IR) remains largely unexplored. } 
   {The IR iron lines around 15600\,$\rm{\AA}$, most commonly known for their high magnetic sensitivity, also have a non-zero response to line-of-sight velocity below $\log (\tau)=0.0$. In this paper we aim to tap this potential to explore the possibility of using them to measure sub-surface convective velocities. }
   {By assuming a snapshot of a three-dimensional magnetohydrodynamic simulation to represent the quiet Sun, we investigate how well the iron IR lines can reproduce the LOS velocity in the cube and up to what depth. We use the recently developed spectropolarimetric inversion code SNAPI and discuss the optimal node placements for the retrieval of reliable results from these spectral lines.  }
   {We find that the IR iron lines can measure the convective velocities down to $\log (\tau)=0.5$, below the photosphere, not only at original resolution of the cube but also when degraded with a reasonable spectral and spatial PSF and stray light. Meanwhile, the commonly used Fe~{\sc i} 6300\,\AA{} line pair performs significantly worse.}
   {Our investigation reveals that the IR iron lines can probe the subsurface convection in the solar photosphere. This paper is a first step towards exploiting this diagnostic potential.}


   \maketitle
   
%

\section{Introduction}

Overshooting of the convective elements from below the solar surface manifests as bright granules surrounded by dark intergranular lanes, a characteristic feature of the solar photosphere. Their properties such as size, lifetime, brightness variation and evolution have been long studied and helped to deepen our understanding of the energy and mass transport to the solar surface \citep{1996ASPC..109..155R,1997ApJ...480..406H, 1999A&A...344L..29B, 2002A&A...385.1056B, 2000A&A...358.1109F, 2000A&A...359.1175K, 2002A&A...381..253B, 2002A&A...381..265C, 2002A&A...385.1056B, 2006A&A...458..941K, 2007A&A...476..341K, 2017ApJ...836...40O, 2017ApJ...849....7O, 2017A&A...598A.126L}. For a detailed review on solar convection see \citet{2009LRSP....6....2N}. 

The line-of-sight (LOS) component of the solar convective velocity can be estimated from the Doppler shift of spectral lines using methods such as the bisector analysis, inversions applied to spectroscopic or spectropolarimetric data, or by measuring the Doppler shift of the line core. Local correlation tracking and methods based on neural networks \citep{2017A&A...604A..11A} have been developed to measure the horizontal component.  Measurements of the LOS velocity using ground and space-based instruments show a general trend of decreasing upflow and downflow with increasing atmospheric heights, consistent with overshooting convection. Over the years, a range of values for their magnitude, have been reported in the literature. For example, \citet{1999A&A...344L..29B}, using THEMIS observations, measured the mean granular velocity to vary between -171\,m/s to -93\,m/s and in the cell boundaries 57\,m/s to 81\,m/s. \cite{2000A&A...359.1175K} measured 1-2\,km/s RMS velocity in the granules. Height stratified convective velocities using multi-line inversions employing multi-component model atmospheres can be found in \citet{2000A&A...358.1109F, 2002A&A...385.1056B, 2006A&A...458..941K}. Recently, \cite{2017ApJ...836...40O} measured amplitudes of upflow and downflow velocities ranging from -3.0km/s to +3.0km/s, and compared them with numerical simulations.  

Spectral lines in the visible range, to name a few, Fe~{\sc i} 5576.1\,\AA, Fe~{\sc ii} 5234\,\AA, the Fe~{\sc i} lines observed by \textit{Hinode} Solar Optical Telescope at 6301\,\AA{} and 6302\,\AA, and Fe~{\sc i} 5250\,\AA, are commonly used for convective velocity measurements. These lines have a typical height of formation between 200\,km -- 400\,km above the solar photosphere. \cite{1999A&A...344L..29B, 2002A&A...381..253B} used the C~{\sc i} 5380.3\,\AA{} line that probes lower heights (roughly 60 km above the photosphere) to study convection at the granules and the cell boundaries. However, probing the solar convection at infrared wavelengths has so far remained unexplored. The well known Fe~{\sc i} lines in the infrared at $15600\,\rm{\AA}$ are known to have a low formation height \citep[e.g.]{2016A&A...596A...2B} and this property of the infrared lines has been exploited to probe deep into the solar photosphere, in the sunspot penumbrae, by \cite{2016A&A...596A...2B}, but also in the quiet Sun \cite[]{2007A&A...469L..39M, 2008A&A...477..953M, 2016A&A...596A...6L, 2016A&A...596A...5M, 2018A&A...616A.109K}. Also, the mean solar spectrum of these lines was modeled, either to infer atomic parameters \citep{2003A&A...404..749B} or to verify state-of-the-art MHD simulations \citep{2013A&A...557A.109B}. The authors most often discuss the ability of these lines to constrain the atmospheric parameters at $\log (\tau_{5000})=0.0$. However, when we examine their response functions, they also sample layers below $\log (\tau_{5000})=0.0$ (discussed in Section~\ref{cubeandprof}). Additionally, their high excitation potential make them less sensitive to temperature fluctuations. This leads us to the question, if is it possible to extract reliable  atmospheric parameters from layers below the photosphere, i.e. at  $\log (\tau_{5000})>0$. If yes, then what is an optimal setup required for the inversions? Exploring this is one of the main aims of the present paper, with a particular focus on the convective velocity. 

We synthesize the Stokes profiles from a three-dimensional magneto hydrodynamic (3D MHD) cube representative of the quiet Sun atmosphere and perform state-of-the-art inversions using the newly developed SNAPI (Spectropolarimetric NLTE Analytically-Powered Inversion) code \citep{2017A&A...601A.100M, 2018A&A...617A..24M}. By experimenting with the placement of nodes, we investigate which are the deepest layers in the solar atmosphere we can reliably diagnose with the infrared lines of iron at $15600\rm{\,\AA}$. We then repeat the process for the widely used iron line pair at 6300\,$\rm{\AA}$, and discuss the differences in the retrieved velocities. The whole procedure is then repeated by degrading the data enough to resemble the observations from the new and upgraded version of the GREGOR Infrared Spectrograph \citep[GRIS;][]{2012AN....333..872C} called GRIS+. Such a study gives us an estimate on the extent of spatial detail we can hope to see from the inversion of this infrared spectral region using GRIS+. We  present a detailed investigation with different levels of degradation. We compare the inverted atmosphere with the original MHD atmosphere and this gives us an idea on the reliability of our diagnostics and helps us determine the optimal setup for the inversion of the actual telescope data.

Section~\ref{cubeandprof} describes the details of the MHD cube and the Stokes profile synthesis. The inversion of synthetic data with different degrees of degradation is explained in Section~\ref{sc:inv}, where the results from the infrared lines are compared with those from the Fe~{\sc i} 6300\,\AA{} line pair. Our conclusions are outlined in Section~\ref{sec:conclusions}. 

   \section{MHD cube and profile synthesis}
   \label{cubeandprof}
   
   For the purpose of this study, we consider a 3D MHD snapshot to be an example of `true' solar atmosphere. Note that it very well might not be the case, as we are currently not sure how well state-of-the-art simulations describe the Sun. That discussion, however, is beyond the scope of this paper. What matters to us is that the snapshot is probably a reasonable imitation of the real atmosphere as most of its observed properties can be reproduced \citep[e.g.][]{2008A&A...484L..17D}. Now, by synthesizing the spectra from the cube and applying instrumental effects to it, we aim to achieve data of similar quality to real observations. If we apply our diagnostics technique (inversion, in this case) to this data, we will retrieve the physical properties of the atmosphere. There is one big difference with respect to the real observations though: we know how the original atmosphere looks, and we can compare the inferred atmosphere with the original one. Thus we are able to test the reliability of our inversion. We focus on the following two aspects:
   \begin{enumerate}
       \item \textit{Retrieving information from deep layers}: since the infrared lines of iron, we are interested in, lie in the spectral region where the $H^-$ opacity reaches minimum, we expect the lines to be more sensitive to the deep layers (i.e. deeper than $\log\tau=0$). By inverting synthetic data and comparing inferred and original atmospheres we can explicitly see up to what depth the two agree  and thus conclude how deep we can probe by using this spectral window. Furthermore, we perform direct comparison with the iron line pair at 6300$\,\rm{\AA}$, i.e. we investigate which line pair recovers the original atmosphere to a higher degree. 
       \item \textit{Seeing effects}: the images of the Sun we obtain from the ground-based telescopes are blurred because of the turbulence in the Earth's atmosphere. This is in addition to the diffraction at the telescope aperture. Adaptive optics and state-of-the art image reconstruction techniques such as Multi-Object Multi-Frame Blind Deconvolution \citep[MOMFBD][]{2005SoPh..228..191V} eliminate this blurring by a certain amount, however, the images are still not infinitely sharp. This means that there still remains spatial structures which cannot be retrieved. To analyze the influence of this effect on the inferred atmosphere, we convolve the synthetic data with a theoretical telescope point spread function (PSF), and then perform the inversion on this degraded, simulated spectropolarimetric dataset. Again, we compare the inverted and the original atmospheres to estimate the amount of information that is lost due to this degradation. We stress that this process is not identical to simply smearing the original MHD datacube with telescope PSF. This is because the relationship between the parameters we want to infer and the observables (Stokes vector) is highly nonlinear. 
   \end{enumerate}
   
   This numerical experiment allows us to characterize the reliability of our inversion and estimate the amount of information we are able to retrieve. In addition, we can prepare an optimal  setup for the inversion of these iron lines, which are often used to observe a variety of solar targets at the GREGOR telescope \citep{2012AN....333..796S} using GRIS \citep{2012AN....333..872C} and are also well suited for the upcoming instruments and facilities like the GRIS+ at GREGOR, European Solar Telescope \citep[EST,][]{2013hsa7.conf..808C} and the Daniel K. Inouye Solar Telescope \citep[DKIST,][]{2014SPIE.9147E..07E}. 

  An additional caveat is in place: here we assume that our synthesis procedure applied to the MHD simulation accurately depicts the way observed Stokes profiles are generated. This might not be the case due to a number of additional effects:
   \begin{itemize}
    \item Possible departures for Fe I populations from LTE due to, e.g. radiative overionisation. This effect is important for the temperature inference but much less so for the velocity.
    \item Presence of mixed atmospheres in the observed pixel (we address this in the subsection \ref{ssection:psf}).
    \item Inaccuracies in the atomic parameters and/or background opacities. 
    \item Presence of systematics in the observed data (cross talk, fringes, etc.).
   \end{itemize}
  In the following, we assume that we can obtain atomic parameters and continuum opacities that are accurate enough, and that our data is properly calibrated, but it is important to stress that in the applications to an actual telescope data this might not be the case.
   
   \subsection{MHD cube description}
   
   The 3D MHD cube used here is a non-gray atmosphere simulated from the MURaM code \citep{2005a&a...429..335v}. The details on the numerical setup of this simulation are described in \citet{2014a&a...568a..13r}. The cube has a horizontal resolution of 20.83\,km and spans 6\,Mm. In the vertical direction, the cube extends  1.4\,Mm with a resolution of 14\,km. The cube was generated by introducing a uniform vertical magnetic field of $B_z=50$\,G into a hydrodynamic atmosphere. This cube is the same as the one used in \citet{2017A&A...608A.111S}. The maps of temperature, LOS velocity, and the magnetic field strength at $\log (\tau_{5000})=0$ can be found in that paper.

\begin{table*}
\label{atomic_table}
\centering
\caption{Details of the infrared lines}
\begin{tabular}{cccccccc}
\hline
Wavelength\,(\AA)& $J_l$ & $J_u$ &log($gf$) & $\chi_{e}$\,(eV) & g$_{\rm eff}$ & $\sigma$ & $\alpha$\\
\hline
15645.02 & 2.0 & 2.0 & -0.65 & 6.31 & 2.09 & 1035 & 0.291 \\	
15648.52 & 1.0 & 1.0 & -0.67 & 5.43 & 3.00 & 542 & 0.228 \\
15652.87 & 5.0 & 4.0 & -0.04 & 6.25 & 1.53 & 751 & 0.259 \\
15662.02 & 5.0 & 5.0 &  0.19 & 5.83 & 1.48 & 658 & 0.239 \\
15665.25 & 1.0 & 5.0 & -0.42 & 5.98 & 0.67 & 689 & 0.234 \\
\end{tabular}
\tablefoot{The columns indicate wavelength, total angular momentum quantum number of the lower ($J_l$) and upper levels ($J_u$), the oscillator strength log$(gf)$, the lower level excitation potential in e.v. ($\chi_e$), Land\'{e} g-factor (g$_{\rm eff}$) and $\sigma$ and $\alpha$ coefficients for collisional broadening computation, respectively.}

\end{table*}

   \subsection{Spectrum synthesis and post processing}
   
   The MHD cube is essentially a 3D array of temperature, gas pressure, magnetic field and velocity vectors on a specified geometrical grid. These quantities, together with the atomic models, are fed into the SNAPI synthesis/inversion code, to synthesize the Stokes spectra of five neutral iron lines in the spectral region between 15640 to 15670\,$\AA$, with 20\,m$\AA$ sampling. The details of the lines are given in table\,\ref{atomic_table}. Their line strengths and Land\'{e} factors differ significantly offering an added advantage, because spectral lines of different strength (opacity), probe different depths in the atmosphere such that the full spectral window provides us with a larger vertical span compared to just one spectral line. Different magnetic sensitivity is crucial in decreasing degeneracy in the inferred parameters (more on this in the following subsection). Since these lines are formed quite deep in the atmosphere, where collisions dominate, we have used the assumption of local thermodynamic equilibrium (LTE), which accelerates the profile synthesis and the inversion. To calculate collisional broadening, we use the approach from \citet{1997MNRAS.290..102B}, and we provide the coefficients we used in the table.
   
   The next step is post-processing the data such that it resembles the real observations. We consider three main effects: spectral smearing in the spectrograph, spatial smearing due to the finite spatial resolution, and the presence of photon noise. Accordingly, we prepare three different versions of the dataset for the infrared lines:
   \begin{itemize}
       \item \textit{Version 0} (v0): The original data without any instrumental effects.
       \item \textit{Version 1} (v1): The original data, spectrally convolved with a Gaussian filter with Full width at half maximum (FWHM) of 150\,m$\AA$ and then resampled to a spectral resolution of $60$\,m$\AA$. To this we added a wavelength independent Gaussian noise equal to $3\times 10^{-4}$ of the quiet Sun continuum intensity. This is the level of noise GRIS+ instrument is expected to deliver when observing the quiet Sun. 
       \item \textit{Version 2} (v2): Same as version 1, except that we now also apply spatial smearing. We assume a Gaussian spatial PSF with FWHM of $0.26^{\arcsec}$ which corresponds approximately  to 200\,km on the solar surface or ten pixels in the original MHD simulation. In addition, we consider another component of PSF, which is also a Gaussian but much broader (FWHM = $2^{\arcsec}$), which mimics the so called straylight. The straylight is a consequence of the broad wings of the PSF and is very hard to correct for using the MOMFBD or similar image restoration techniques. We assume different amounts of straylight (10\%--30\%), and refer to these versions as v2.1, v2.2, and v2.3, respectively. For inversion purposes, these data cubes have also been rebinned to $3\times3$ bins, which correspond to spatial sampling of $0.085^{\arcsec}$ in the plane of the sky.
       
   \end{itemize}
   
\begin{figure}
       \centering
       \includegraphics[width=0.24\textwidth]{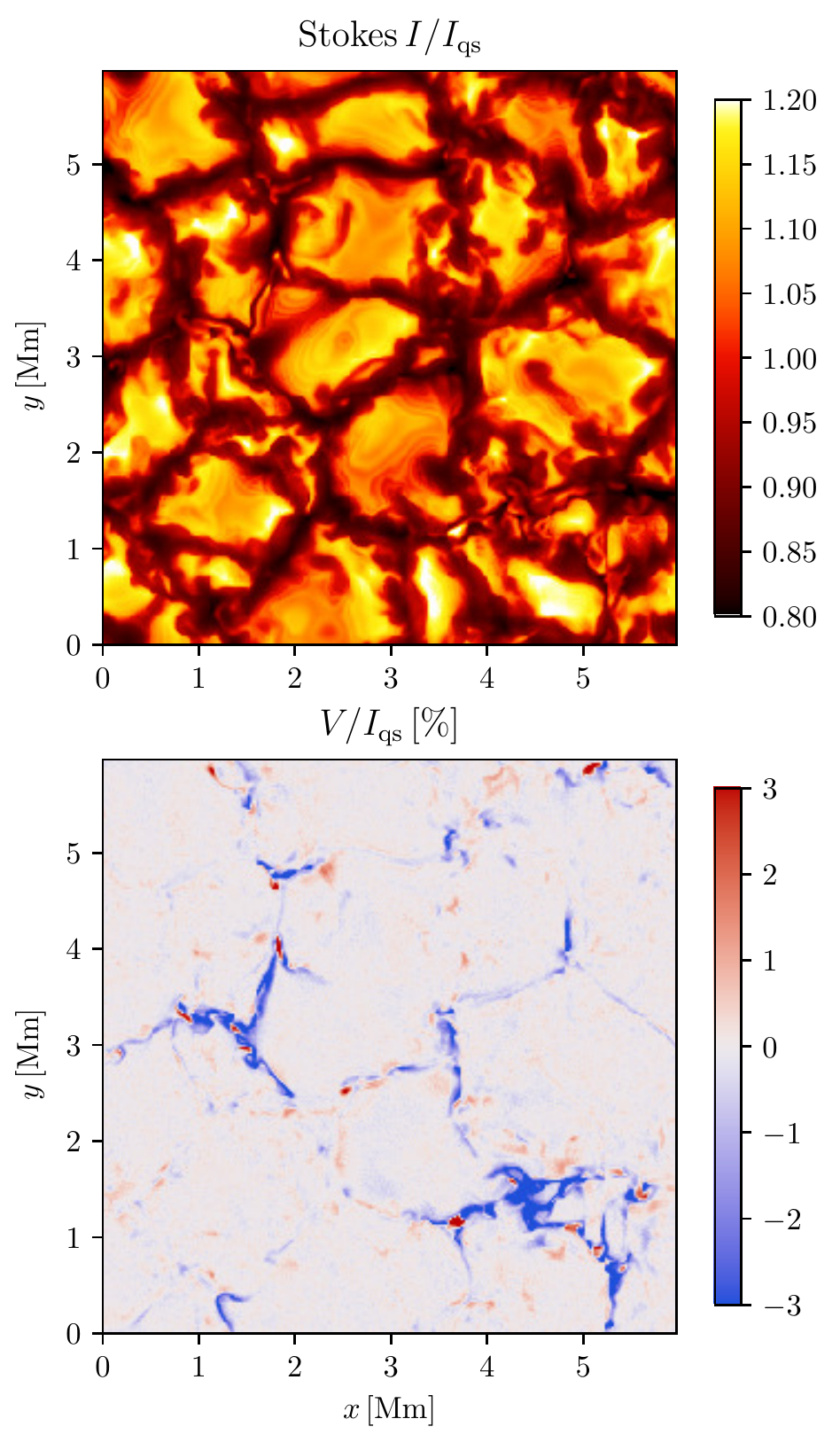}
       \includegraphics[width=0.24\textwidth]{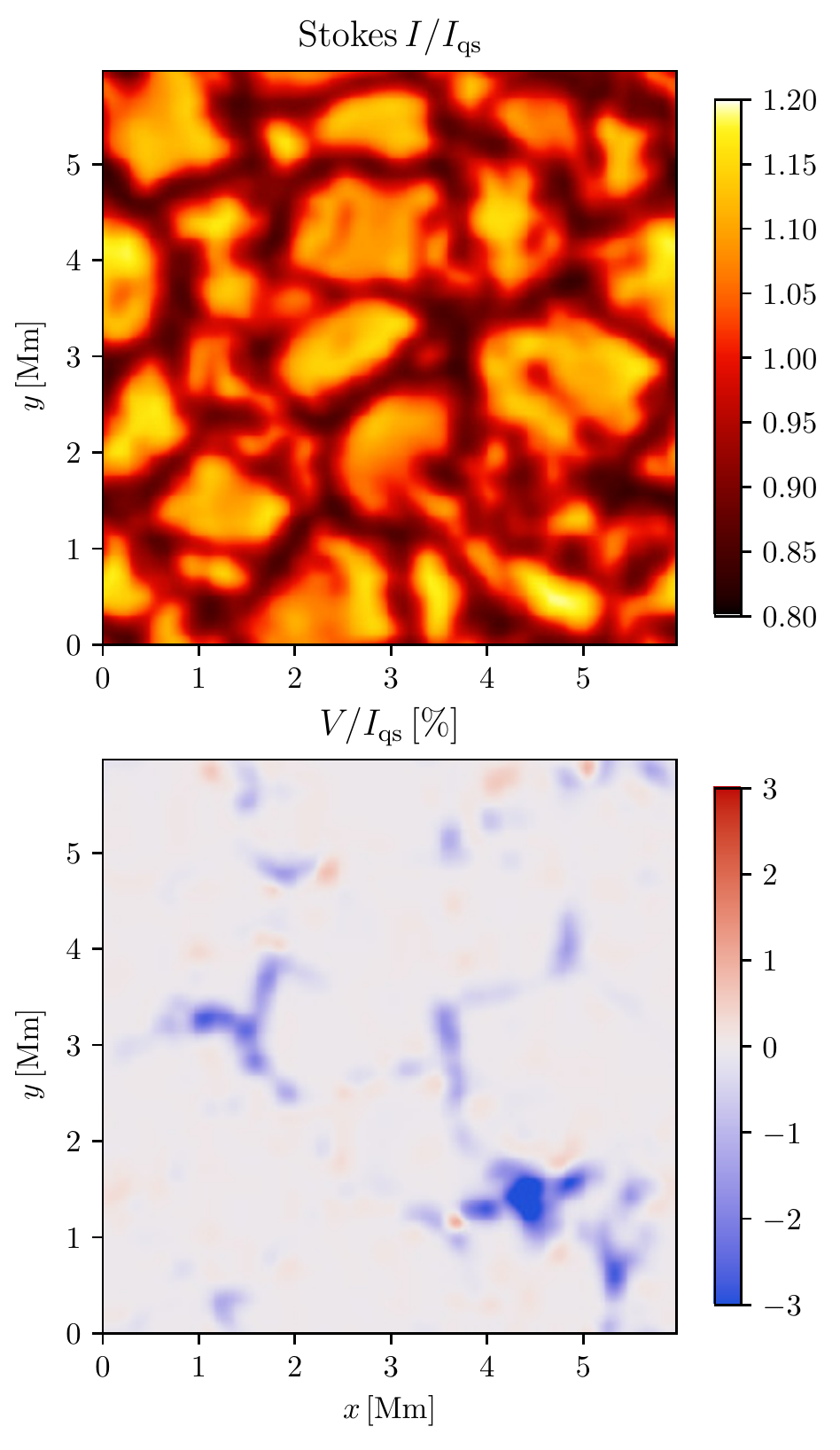}
       \caption{Original (left), and convolved (right) synthetic images, resulting from the MURaM cube. Upper panels show intensity in the continuum and lower ones the circular polarization in the wing of 15662$\,\AA$ line.}
       \label{example_data}
   \end{figure}

   An example of the v1 and v2 synthesized data is shown in Fig.\,\ref{example_data}. The loss of spatial detail after the convolution with $0.26^{\arcsec}$ wide PSF is evident, although this would be a rather good resolution for a ground based telescope so far in the infrared. Our goal now is to analyze this data, treating them like real observations.
   
   \section{Inversions}
   \label{sc:inv}
   Spectropolarimetric inversion is the process of fitting a model atmosphere to the observed Stokes spectrum. The model atmosphere (or the corrections to it), are usually parametrized using nodes, pre-chosen points in depth, where the atmosphere is allowed to vary while the values in between the nodes are interpolated (usually using splines).  This parametrization reduces the number of free parameters and keeps the inferred atmospheres relatively simple. Inversions allow us to infer the spatial distribution and height stratification of the temperature, LOS velocity and magnetic field in the observed patch of the solar surface, making it one of the preferred tools for quantitative interpretation of spectropolarimetric data. For an excellent recent review on spectropolarimetric inversions, see \citet{2016LRSP...13....4D}.
   
   In this section we present the inversions of the synthetic data described above. We again use SNAPI, which is an inversion code intended for NLTE lines, but can also be applied to lines formed in LTE conditions and provides increased flexibility in positioning of nodes for different parameters.
   
   We chose an atmospheric model with five nodes in temperature, four nodes in the LOS velocity and two in magnetic field strength with constant magnetic field inclination and microturbulent velocity. As our focus is mainly on the velocity diagnostics, we chose to fit only the Stokes $I$ and $V$ profiles (besides, the linear polarization still exhibits relatively high level of noise). Adding information from Stokes $V$ helps us determine magnetic field strength, and thus avoid possible degeneracy between the magnetic field and velocity. The node positions are given in Table~\ref{tab:nodes}. Note that $\tau$ refers to the continuum optical depth at 5000\,$\AA$, so the deepest nodes are lying underneath what we usually refer to as the photosphere ($\log\tau_{5000}=0$). We stress that optimal choice of nodes (their number and position) for the parametrization of the atmosphere is a matter of constant discussion in the spectropolarimetric community. While this has often been referred to as an art \citep{2016LRSP...13....4D}, a detailed systematic investigation using model fitting is definitely lacking. 
   
   \begin{table}[]
   \caption{Caption}
       \centering
       \begin{tabular}{c c}
        Parameter & Node positions [$\log\tau_{5000}$]\\ \hline   
        Temperature   &  -3.4, -2.0, -0.8, 0.0, 0.5 \\
        LOS velocity & -2.5, -1.5, -0.5, 0.5 \\
        Magnetic field strength & -1.5, 0.3 \\
        Microturbulent velocity & const \\
        Magnetic field inclination & const
       \end{tabular}
    
       \label{tab:nodes}
   \end{table}
   
\begin{figure}[htbp]
       \centering
       \includegraphics[width=0.5\textwidth]{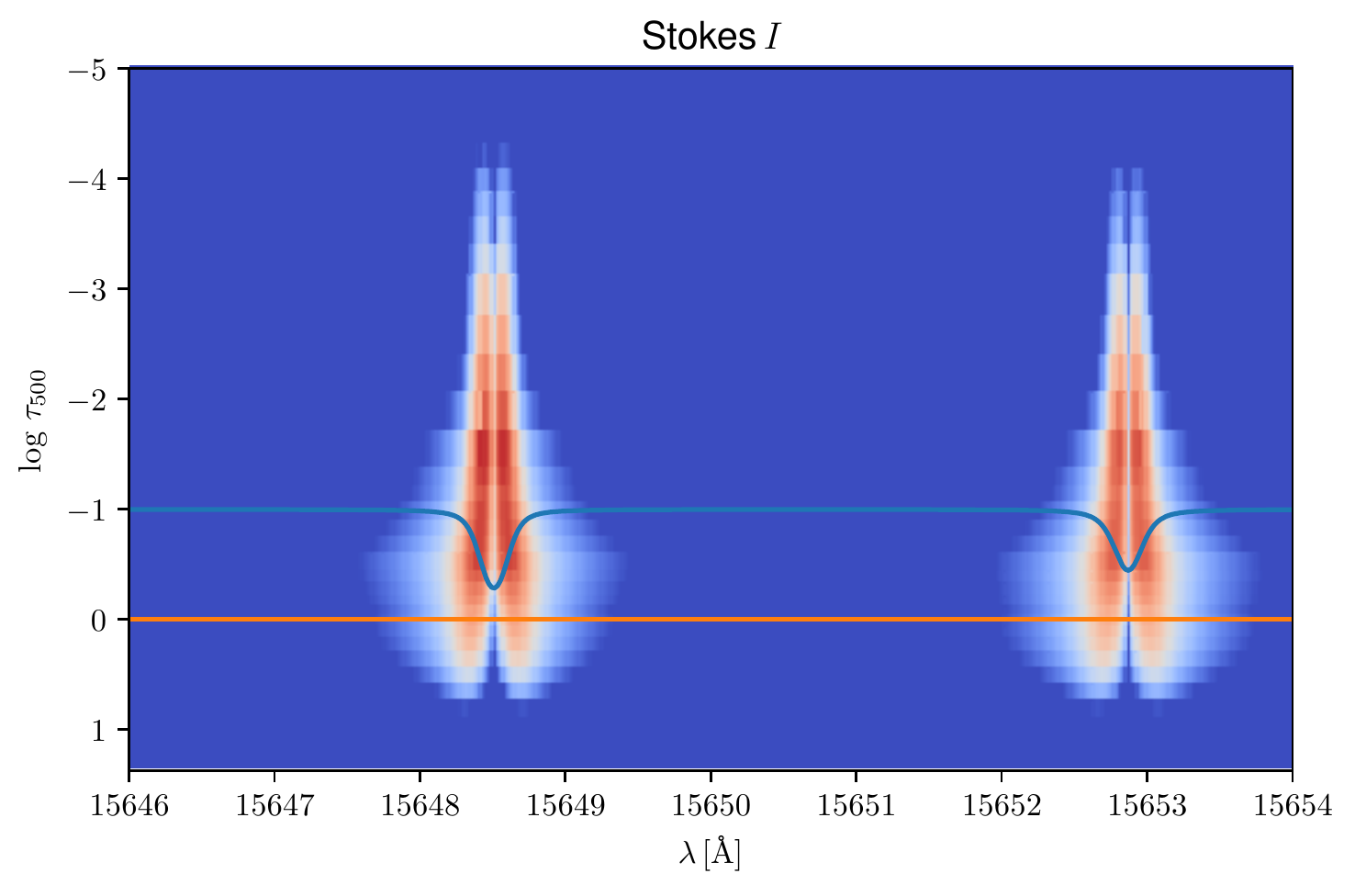}
       \includegraphics[width=0.5\textwidth]{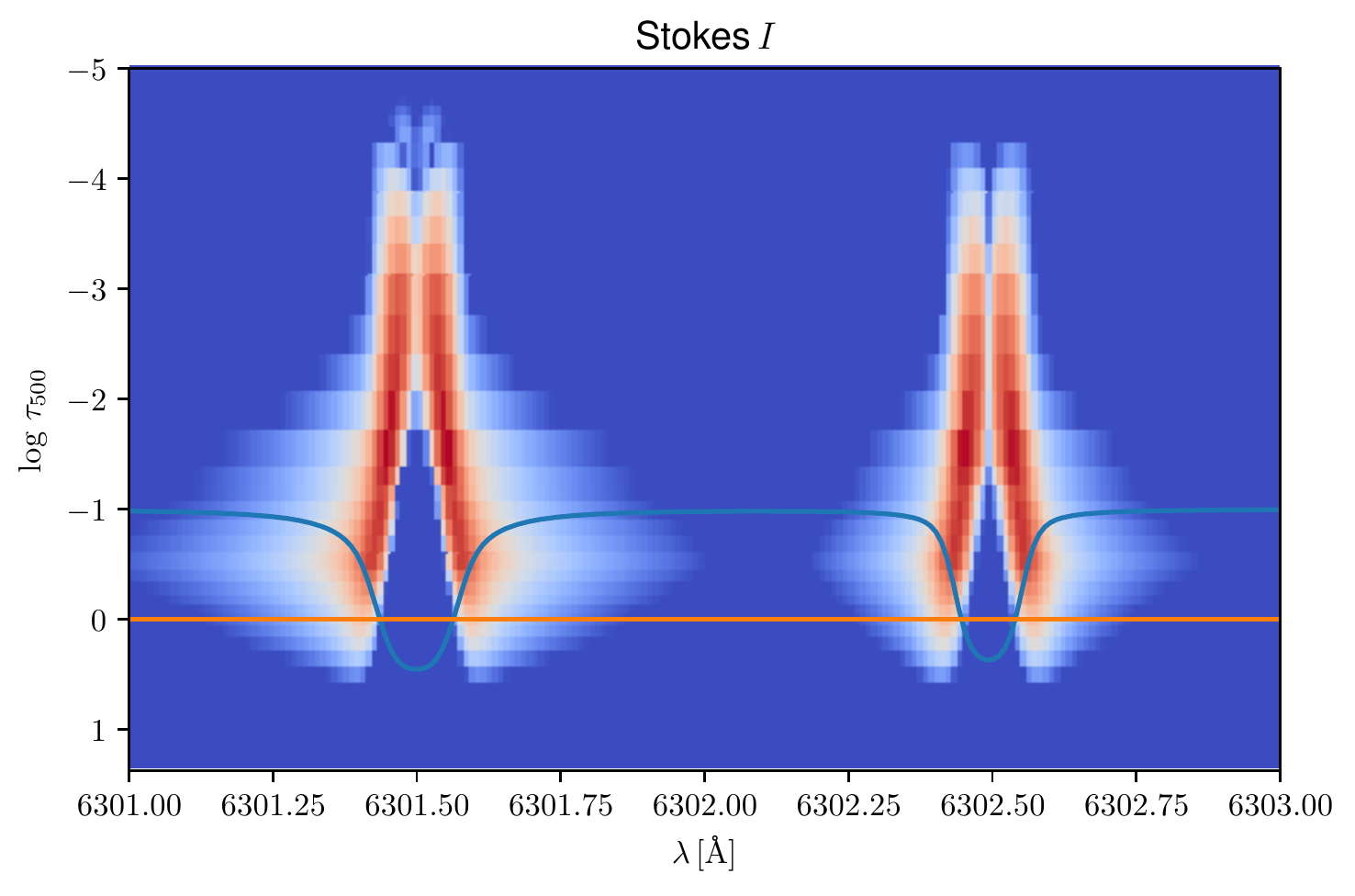}
       \caption{Response functions of Stokes $I$ to the line of sight velocity, for the two redmost infrared iron lines (up), and for the iron pair at 6300\,\AA (down). Both response functions are normalized with respect to their maximum value. Blue line outlines the shape and the position of the spectral lines and red one gives emphasizes $\log\tau_{5000}$ surface.}
       \label{rfs}
   \end{figure}

   This node positioning has been particularly chosen to probe LOS velocities in the range $-1.5<\log\tau<0.5$. The uppermost velocity node ($\log\tau=-2.5$) basically sets the derivative of the velocity in the node at $\log\tau=-1.5$. The sensitivity of the chosen infrared lines to layers below $\log\tau_{5000}$ can be seen from the response functions of Stokes $I$ to the LOS velocity (Fig.\,\ref{rfs}). Response functions \citep{1977A&A....56..111L, 1994A&A...283..129R} are a measure of how the emergent spectrum responds to perturbation of different physical quantities at different depths, and these particular ones are calculated from using FALC \citep{1993ApJ...406..319F} semi-empirical model atmosphere. Response functions provide information on the diagnostic potential of the lines in the selected wavelength region and their calculation is necessary for the inversion procedure. From the figure, it is clear that radiation from the chosen spectral lines is sensitive to velocity perturbations from below the photosphere. Though the visible lines at 6300\,\AA{} also have non-zero response for $\log\tau>0.0$, the responses from the infrared lines seem to be somewhat stronger.
   
    Note that we include microturbulent velocity in the model used for inversion, although the original MHD cube contains no such quantity. Microturbulent velocity was introduced in the early solar atmospheric modeling where it was used in order to explain the spectrum of the spatially unresolved Sun using 1D models. Originally, microturbulent velocity accounted for all the unresolved velocities, in all three coordinates. With today's state-of-the art instruments, we can resolve velocities in $(x,y)$ plane (i.e. in the plane of the sky) rather well. Still, we use the microturbulent velocity in inversions, in order to account for the complicated depth structure in velocity which cannot be fully resolved by our simple, node-based model. Finally, we stress that the term microturbulent is confusing as there is no consensus on the relationship between these, ad-hoc introduced, velocities and turbulent processes in solar plasma. 

  Note that, as is commonly done, we eliminate gas pressure as a free parameter by using the assumption of hydrostatic equilibrium. To do this, one can either assume an upper boundary condition for the pressure, or make additional assumptions. SNAPI code relies on the approach from \citet{1989isa2.book.....B}, where the atmosphere is assumed to be exponentially stratified above the uppermost point, and then the pressure in that point is obtained by iteration.
   
   The synthetic spectra are inverted using a multi-cycle approach for the spatial regularization. That is, after every 10 fitting iterations we apply 2D median and Gaussian filters in the $(x,y)$ plane, to each of the model parameters, which are in the case of SNAPI, values in the pre-determined nodes in optical depth \citep[for more details on node approach, see, e.g.][]{2018A&A...617A..24M, 2018arXiv181008441D}. This ensures the smoothness of the parameter maps and helps avoid local minima. The process is repeated until both average $\chi^2$ and the maximum $\chi^2$ in all the pixels are not changing substantially (typically 6-7 cycles, or $\approx 100$ iterations in total). After obtaining the converged solution, we:
   \begin{enumerate}
       \item Compare the observed and fitted spectra to make sure that the agreement is satisfactory. 
       \item Compare the inferred and the original atmospheres and check if we are able to reliably retrieve velocities at the optical depths we are interested in. 
   \end{enumerate}
   
 \subsection{Inversion of the data with no telescope PSF}
 \label{ssection:no_psf}
 
\begin{figure}[htbp]
     \centering
     \includegraphics[width=0.5\textwidth]{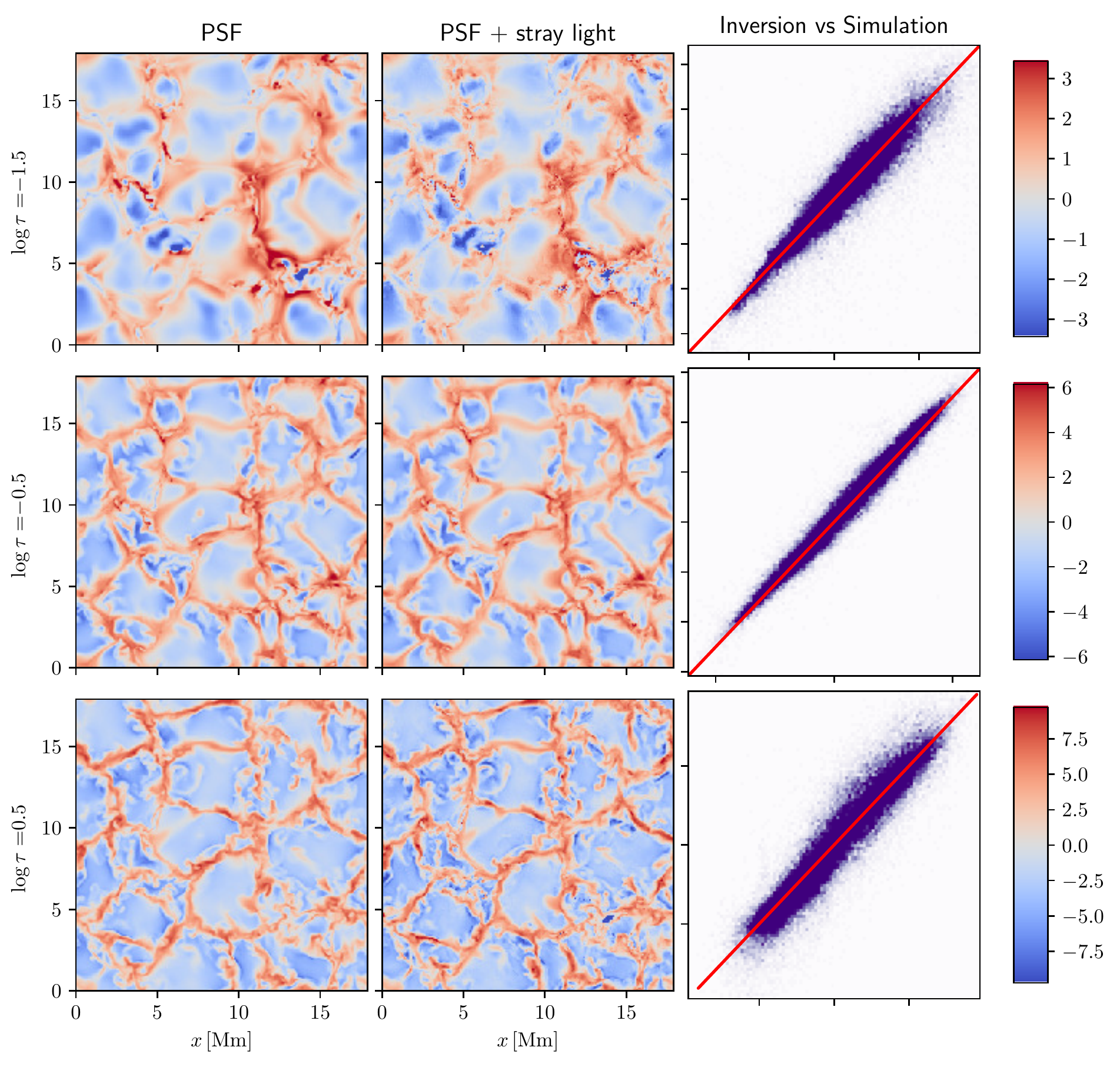}
     \caption{Comparison between line of sight velocities in the MHD simulation and as returned by inversion code, from the undegraded synthetic data, at three different depths in the atmosphere ($\log\tau = -1.5, -0.5, 0.5$). In all the figures, the LOS velocities are indicated in km/s.}
     \label{fig:v_comparison_0}
 \end{figure}

 As a first test of our inversion setup we first invert the original data, without any noise and with the original sampling (that is, v0 data), and compare the original MHD atmosphere with the inverted one. From now on we will only show comparison between velocities.  Note that in the original cube we have the physical parameters on $(x,y,z)$ grid. Inversions, however, can only retrieve parameters on $(x,y,\tau)$ grid, due to the nature of radiative transfer process (from now on we use $\tau$ to denote continuum optical depth at $5000\,\rm{\AA}$). Therefore, we must change $z$ coordinate in the original MHD cube to $\tau$. SNAPI code does this automatically during the synthesis. We chose three different depths to compare at: $\log\tau=0.5$ (deep photosphere), $\log\tau=-0.5$ (a layer where commonly used lines in the optical domains are sensitive to velocity) and $\log\tau=-1.5$ (upper photosphere). Fig.\,\ref{fig:v_comparison_0} shows agreement between velocities in these three atmospheric depths. The velocity map in the upper layers ($\log\tau=-1.5$) is somewhat more noisy, while the velocities in the deep photosphere seem to show slightly higher values than the original ones. However, from the point of the velocity field distribution, agreement is generally excellent at each of the three depths, which shows that this spectral region carries enough information to reliably retrieve velocities at different depths, assuming the observations have very high spectral resolution and very low noise. 
 
\begin{figure*}[htbp]
     \centering
     \includegraphics[width=1.0\textwidth]{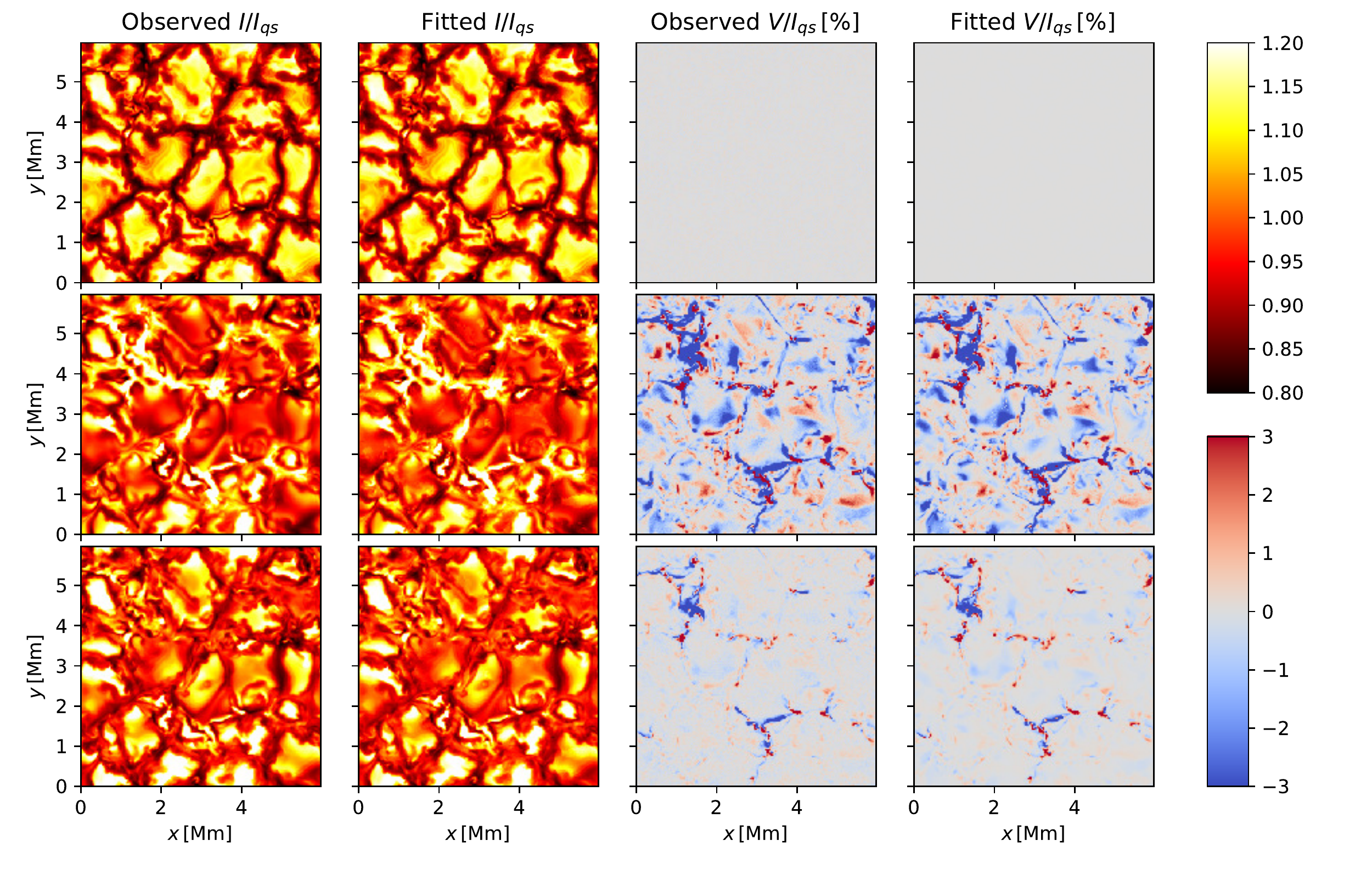}
     \caption{Observed (first and third column), and best fit (second and fourth column) images at wavelengths corresponding to the continuum (topmost), and the line centers of the two redmost IR spectral lines considered in this study. Note that we invert all five lines and just show the two redmost for conciseness. The inverted data is the version 1 data (convolved with spectra PSF and with noise added). Stokes $V$ images correspond to wavelengths shifted by $100\,\rm{m\AA}$ to the red.}
     \label{fig:spectra_comparison_1}
 \end{figure*}

 We then show the results from the inversion of synthesized spectra which has been convolved with spectrograph profile and with added photon noise, but without spatial convolution (i.e. v1 data) and qualitatively compare the original and the inferred atmospheres. As a first check, we assess the goodness of the fit to the synthetic data by comparing the observed and fitted Stokes $I/I_{\rm qs}$ and $V/I_{\rm qs}$ at the continuum and at the core of the two redmost spectral lines ( Fig.\,\ref{fig:spectra_comparison_1}). Here $I_{\rm qs}$ is the spatially averaged continuum intensity at 15600$\,\rm{\AA}$.
 
 The fit quality in both the Stokes parameters is very good, with the average reduced $\chi^2$ around 10. Note that we fit the synthetic observations of all the five lines, but show only fits for the two for conciseness sake. The fits in the other three lines are equally good. When discussing this particular value of $
 \chi^2$, one must keep in mind that in spectropolarimetric inversions we are working with simplified atmospheres (with parameters defined just at the nodes) and that a noise of $3\times10^{-4}$ is quite low. In our case, we obtain the average agreement between the observed and fitted spectra of around $0.1\%$ which is something one would be extremely satisfied with when fitting an actual data from a telescope. This is, of course, neglecting all the possible systematics in the data, as well as possible inadequacies of the model (inaccurate atomic parameters, etc).
 
\begin{figure}[htbp]
     \centering
     \includegraphics[width=0.5\textwidth]{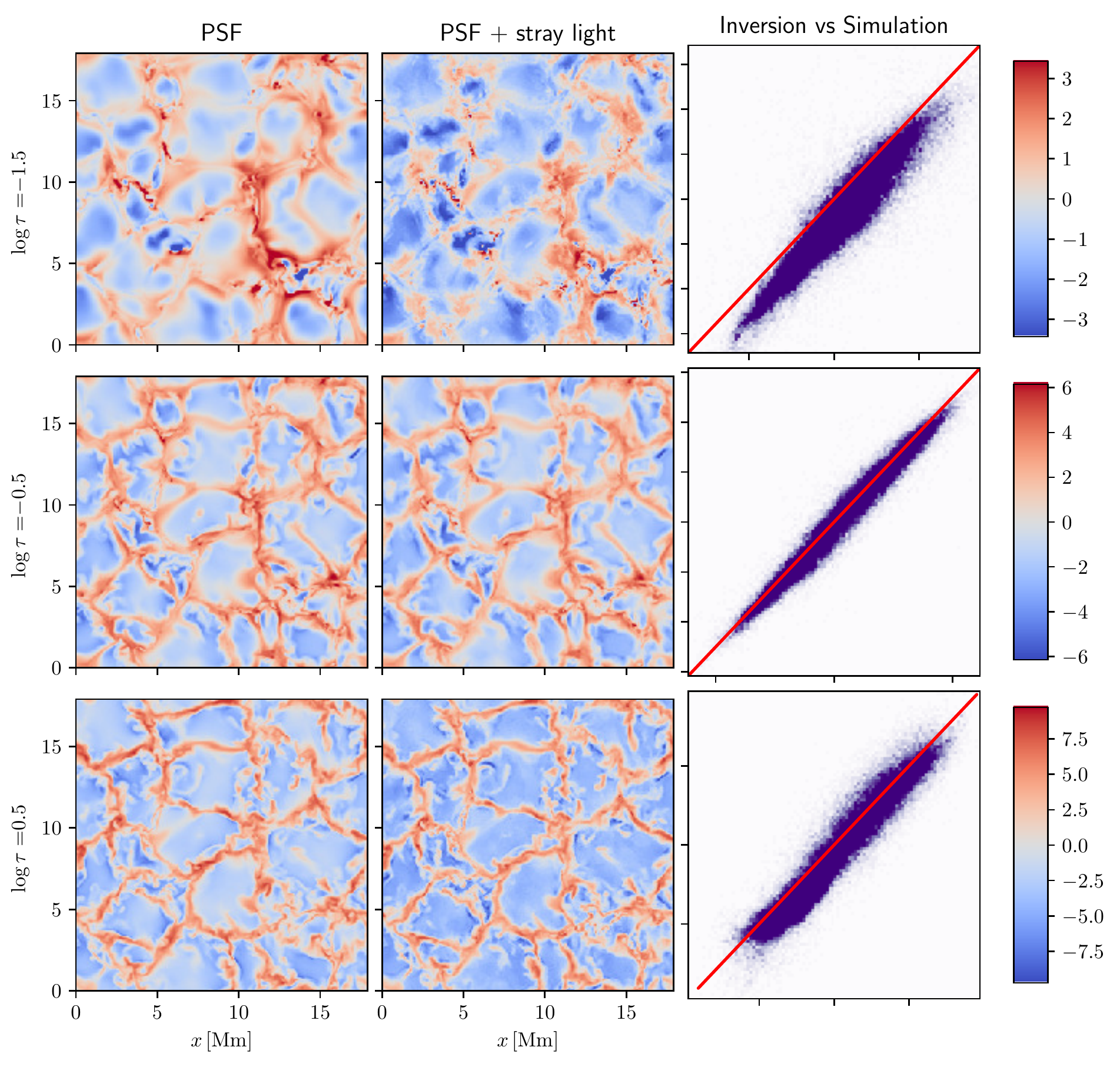}
     \caption{Same as Fig.\,\ref{fig:v_comparison_0}, except the spectra used for inversion have been convolved in wavelength to account for finite spectral resolution, and photon noise has been added.} 
     \label{fig:v_comparison_1}
 \end{figure}

 We show the comparison between the original and inferred velocity structure in Fig.\,\ref{fig:v_comparison_1}. The agreement in the two lowermost depths is still remarkably good and we can conclude that the velocity structure of the atmosphere is well reproduced. In the upper layers ($\log\tau=-1.5$), the maps are slightly more noisy. Our explanation is as follows: convolving and resampling the data in wavelength results in smearing out the sensitivity of the line cores and fewer data points from the line cores are sampled. Since the line cores are the spectral regions which are the most sensitive to the upper atmosphere, the sensitivity decreases. Still, the topology of the velocity field is well retrieved and there are no systematic offsets.
 
 \subsubsection{Probing even deeper layers}
 
\begin{figure}[htbp]
     \centering
     \includegraphics[width=0.5\textwidth]{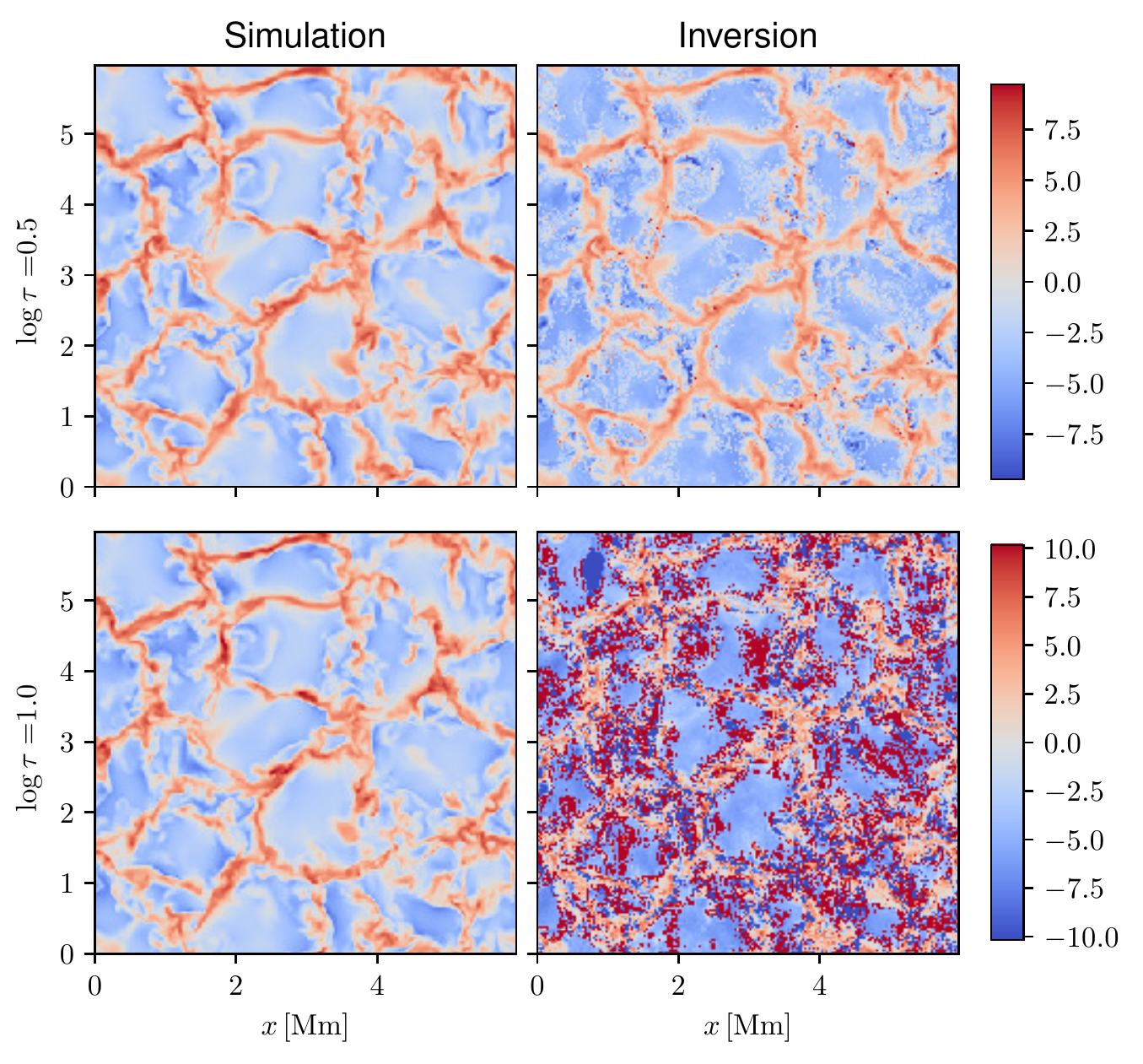}
     \caption{Same as Fig.\,\ref{fig:v_comparison_1}, except with now also invert and compare the plasma velocity at $\log\tau=1$, and we compare the atmospheres at the different optical depths, namely $\log\tau = -0.5, 0.5, 1$.} 
     \label{fig:v_comparison_1a}
 \end{figure}

 In order to test the limits of the sensitivity of those lines, we inverted the whole cube again, adding a node in velocity at $\log\tau=1$. In the semi-empiric FALC atmosphere model, this corresponds to an extra $\approx 30\,\rm{km}$ in depth. If the spectra is sensitive to these depths, we expect the velocities to be well-constrained. Fig.\,\ref{fig:v_comparison_1a} shows that this, however, is not the case. Although the agreement at $\log\tau=0.5$ is still very good, velocity maps at $\log\tau=1$ abound in salt-and-pepper noise and artifacts. This indicates that sensitivity there is significantly lower, and that it does not make sense to try and infer velocities so deep in the atmosphere. We stress that this is not a consequence of simple degeneracy (also known as the cross talk) between the values of velocity in the two lowermost nodes, as, in that case, we would expect both parameter maps to be noisy. 
 
 \subsection{Comparison with Fe {\sc i} 6300\,\AA{} line pair}
 
 The analysis above clearly illustrates the capability of this infrared spectral region to retrieve sub-photospheric (i.e. below $\log\tau_{5000}$) velocities. However, to truly evaluate its advantages and disadvantages with respect to visible lines, we repeated the analysis with the Fe\,6300\,\AA{} line pair. These two spectral lines are widely used for magnetic field and velocity diagnostics \citep[see e.g.][ but also numerous recent works using data from \textit{Hinode} SOT/SP instrument]{1993ApJ...418..928L,1997Natur.389...47W,2001ApJ...547.1130W}.
 
 The data preparation and degradation process is identical, except that we calculate the spectrum with $10\,\rm{m\AA}$ sampling, and then convolve it with $20\,\rm{m\AA}$ FWHM Gaussian spectral PSF. Continuum intensity at 6300\,\AA{} is around 7 times higher than in the previously discussed infrared region, but the spectral sampling is 6 times more dense. This means that the number of photons arriving into our hypothetical detector per spectral bin is approximately the same, assuming the same spatial sampling and transmission. Hence, we degraded the data with the same relative photon noise as in the example above.
 
\begin{figure}[htbp]
     \centering
     \includegraphics[width=0.5\textwidth]{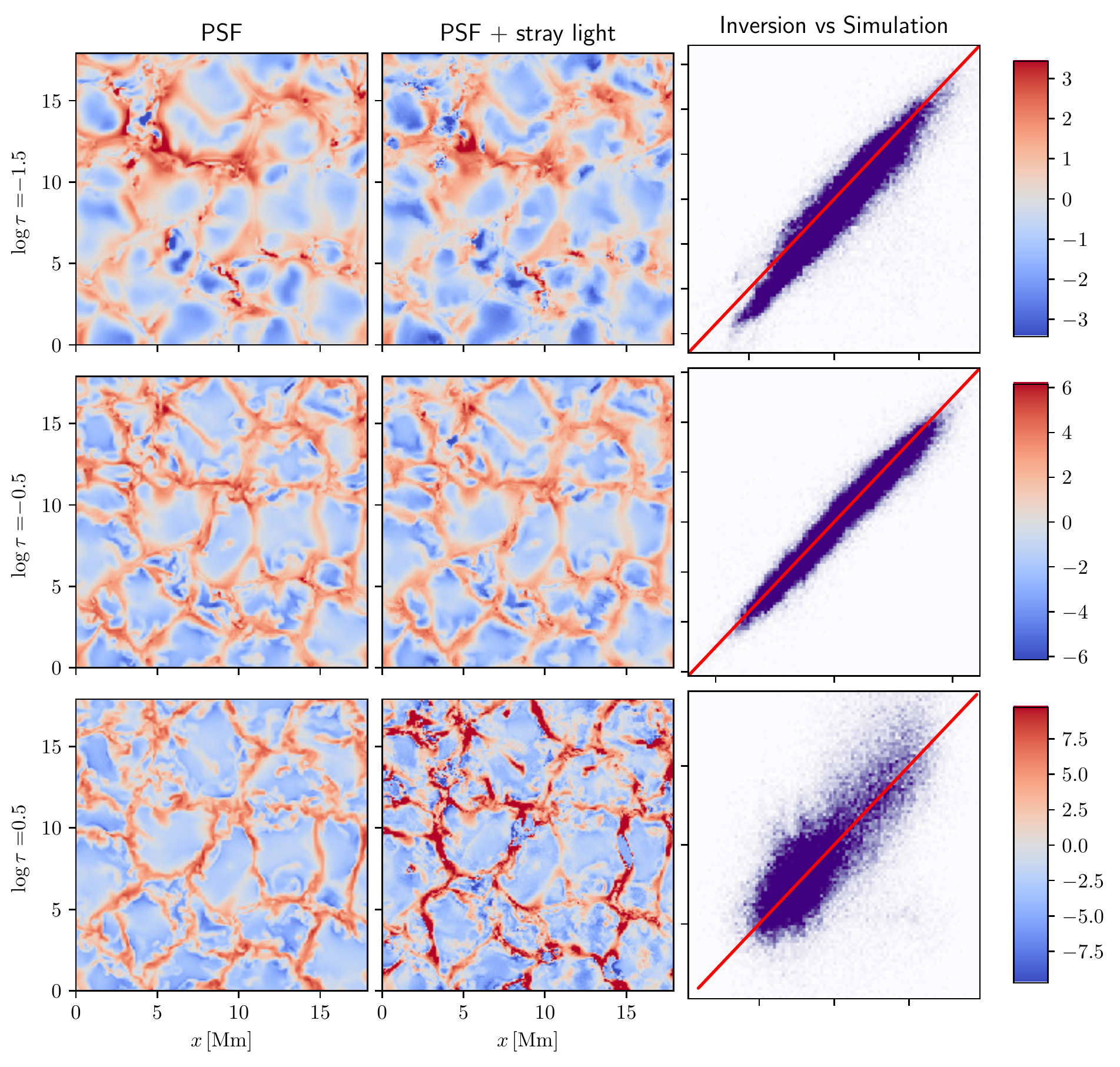}
     \caption{same as Fig.\,\ref{fig:v_comparison_1}, but using the Fe\,6300\,\AA{} line pair.} 
     \label{fig:v_comparison_6300}
 \end{figure}

 To make comparison as straightforward as possible we decided to fit the Fe\,6300\,\AA{} data with the same node configuration as the infrared data (see Table~\ref{tab:nodes} for the exact node configuration). Since the number of nodes is much larger than what is usually used to invert \textit{Hinode} data \citep[e.g.][]{2016A&A...593A..93D}, we took special care to ensure that our retrieved atmospheres are smooth, by employing a multicycle approach where one starts with a smaller number of nodes and then works their way up toward the more complicated models \citep[as is done, in, for example, SIR code by][]{1992ApJ...398..375R}. Now we are able to compare which of the two synthetic data set (infrared or the visible one) constrains the model better. We are focusing on the velocity retrieval. Fig.\,\ref{fig:v_comparison_6300} shows the original and inferred velocities at the same three optical depths as in the previous example. In the upper layers ($\log\tau=-1.5$), the agreement looks slightly better than in the case of infrared lines. In the middle layers, the agreement is equally good. However, in the deep layers, there is large overshoot of the intergranular velocities. Our interpretation is that, because of the low sensitivity of the lines to deep velocities, the inversion code is overestimating the magnitude, in order to get correct derivative of the velocity in the upper layers, while misdiagnosing the actual velocity values at $\log\tau = 0.5$. The better performance of the infrared lines in the deeper layers can also be seen from the scatter plots between the original and the inferred values in the figures \ref{fig:v_comparison_1} and \ref{fig:v_comparison_6300}.
 
\begin{figure}[htbp]
     \centering
     \includegraphics[width=0.5\textwidth]{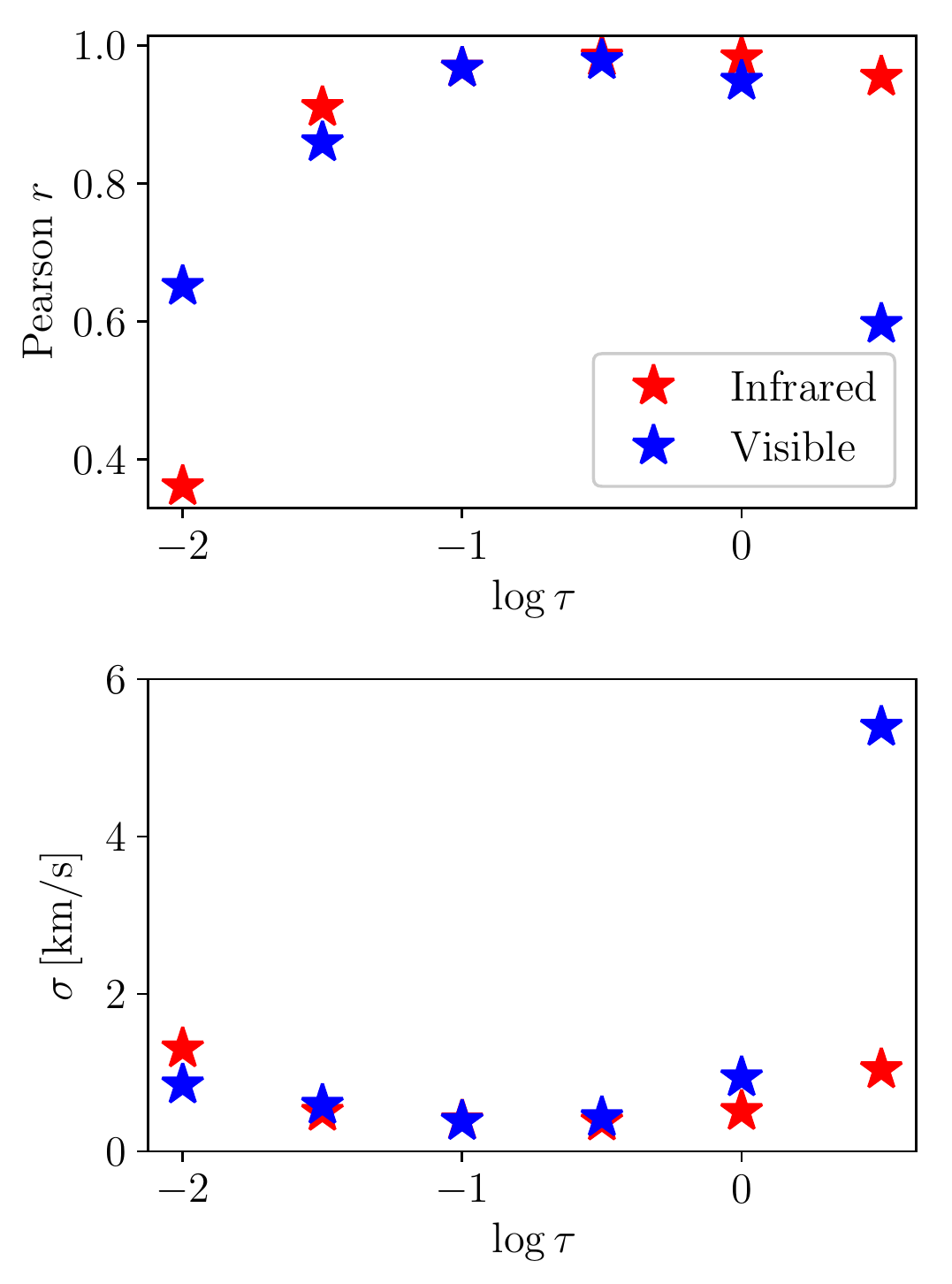}
     \caption{Top: Pearson's correlation coefficient between the velocities in the original MHD cube, and the velocities inferred by spectropolarimetric inversion from the two sets of lines (value closer to unity is better). Bottom: standard deviation of the absolute difference between the original and the inferred velocities (lower values are better).} 
     \label{fig:q_comparison}
 \end{figure}

 While plotting the distributions of the velocities gives us an insight into diagnostic potential, it is beneficial to come up with a quantitative measure of the agreement and compare it between different sets of lines. The simplest is to compare the standard deviation of the absolute difference between the line of sight velocities in original and inferred atmospheres, at several depths. We present such a plot in Fig.\,\ref{fig:q_comparison}. We also plot Pearson's correlation coefficients between the original and inferred velocities at multiple optical depths. Both quantities confirm what we see in the images: infrared lines retrieve better sub-photospheric velocities while in the upper layers both sets of lines give similar results. We note that standard deviation and the correlation coefficients are subject to noise (which increases scatter) and the node configuration (that can systematically change agreement between original and inferred atmospheres). However, these results confirm the prevailing understanding that the infrared iron lines are, in general, better at probing deeper layers of the atmosphere. We did not test agreement in the even higher layers thoroughly, but we note that at $\log\tau=-2$, reliability of infrared lines suddenly drops. This is similar to what happens with visible lines at $\log\tau=0.5$. This suggests that the sensitivity of spectral lines to velocity is confined to a well-defined range of optical depths.
 
 \subsection{Inversion of the spatially smeared data}
 \label{ssection:psf}
 
 In the previous subsections we have discussed the retrieval of the original atmosphere under the implicit assumption that we are able to resolve it spatially. However, we are almost sure that with today's resolving power the part of the solar atmosphere that fits into a resolution element is not homogeneous in the $(x,y)$-plane. That is, we are observing a mixture of 1D atmospheres. We have already illustrated its effect on the observables in Fig.\,\ref{fig:spectra_comparison_1}.
 
 Now we discuss the atmosphere obtained by inverting synthesized spectra which have been subsequently spatially smeared using a Gaussian filter with FWHM equal to $0.26^{\arcsec}$, corresponding to the diffraction limit for GREGOR telescope at 15600\,$\rm{\AA}$. Under good seeing conditions, and after the application of the spectral restoration technique \cite[]{2017A&A...608A..76V}, one can realistically expect to achieve a spatial resolution close to this. 
 
 Inversion of the data convolved with the telescope PSF will result in a 3D atmosphere. We stress that this atmosphere is not identical to the original atmosphere, convolved with the PSF, because spectral synthesis / inversion and PSF application do not commute. Or, said differently: the atmosphere that explains PSF-smeared observations is not the same as the PSF-smeared original atmosphere.

 Before discussing the results we note that there is an another effect that will spatially couple the neighboring atmospheres and that is multidimensional radiative transfer in the presence of scattering. These effects can be safely neglected for the lines we are considering but become very important for scattering dominated lines formed in the upper photosphere and the chromosphere.
 
\begin{figure}[htbp]
     \centering
     \includegraphics[width=0.5\textwidth]{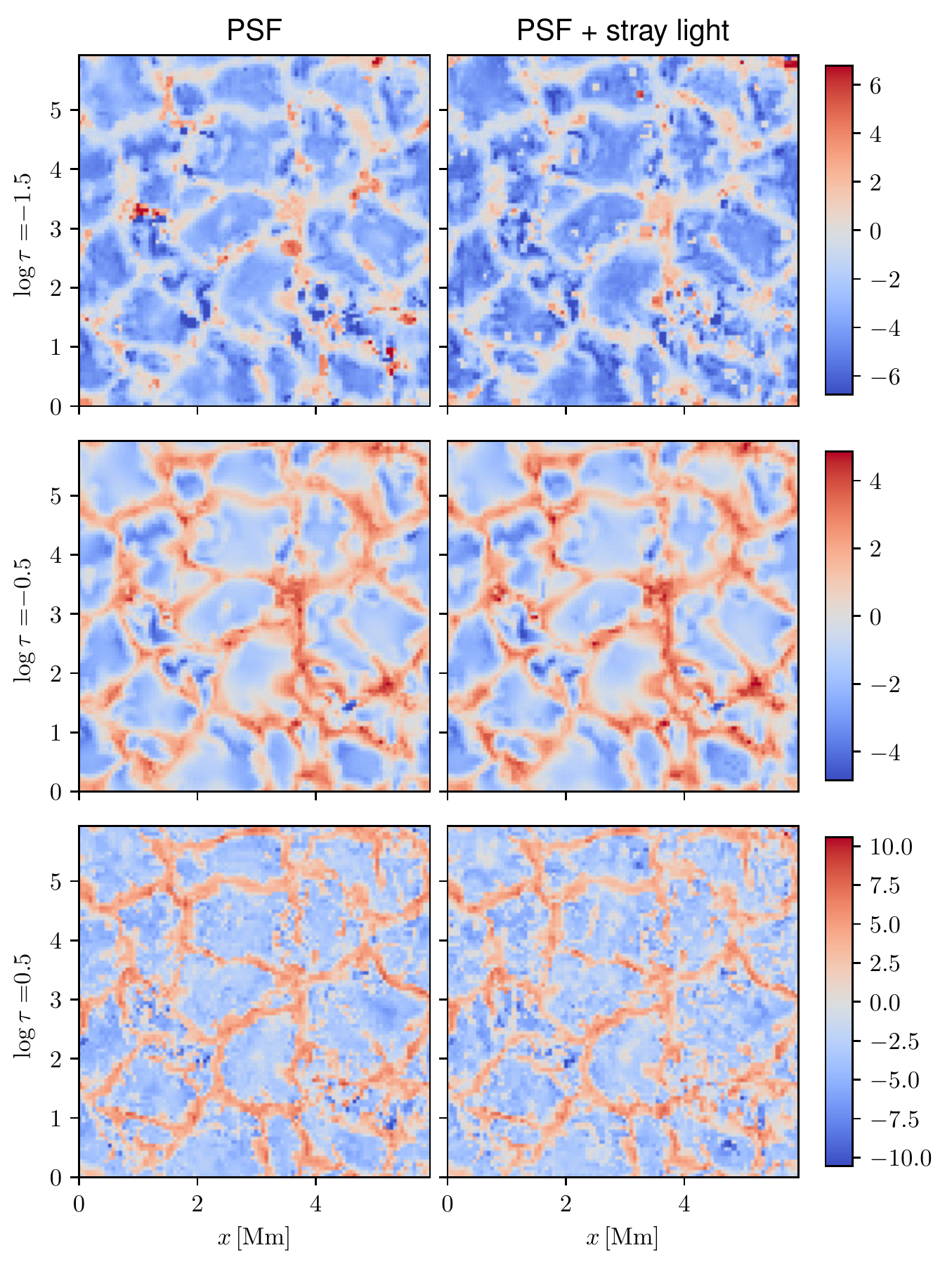}
     \caption{Comparison between the inferred velocities at three optical depths for the data smeared only with narrow PSF (left), and with added stray-light (right).} 
     \label{fig:v_comparison_2}
 \end{figure}

 The fit of the convolved profiles is again excellent. The inferred atmospheres, although, show a significant loss of spatial detail, as expected. While the topology and the velocities are rather well recovered in the middle and lower layers, the discrepancy at $\log\tau_{5000}=-1.5$ is large. Notably, the surface area covered by downflows is much smaller and it also seems that the downflow intensity is lower (left panel of Fig.\,\ref{fig:v_comparison_2}).
 
 
 It is not obvious how to statistically compare the simulated and inverted atmospheres in this case.  We prefer not to convolve the original atmosphere with PSF since, as we have already stressed, PSF can only operate on the observables. For the moment we conclude that the velocities around $\log\tau = -0.5$ and $\log\tau = 0.5$ are well recovered, despite the loss of resolution, and that eventual use of telescopes with larger apertures will allow us to resolve finer structures of the velocity field (see following subsection for more quantitative analysis).

 \subsection{Influence of stray light}
 
 In addition to the relatively narrow and arguably well understood PSF of the telescope, that is in large part compensated for by using adaptive optics and restoration techniques, solar images often suffer from non-negligible stray-light. This stray-light can be understood as relatively spatially flat, wavelength dependent background added to the spectral cube. Its origin is probably in the very wide wings of the PSF which the adaptive optics and image reconstruction cannot compensate for. In our study we modeled it as an additional, broader Gaussian (FWHM = $5^{\arcsec}$). So, when accounting for stray-light we generated our simulated images as:
 \begin{align}
     I^{\rm obs}(x,y,\lambda) = &(1-k) I^{\rm synth}(x,y,\lambda) \circledast G(\sigma_1) \nonumber \\
     &+ k I^{\rm synth}(x,y,\lambda) \circledast G(\sigma_2). 
 \end{align}
 Here $G(\sigma)$ denotes a Gaussian with FWHM equal to $\sigma$, $k$ is the amount of stray-light that we vary from 0 to 30\%, $\sigma_1=0.^{\arcsec}26$ and $\sigma_2=2^{\arcsec}$. The operator $\circledast$ represents convolution. In the first approximation, in the observations of the quiet Sun, the stray-light adds some amount of mean quiet Sun spectrum to each individual pixel, thus lowering the contrast, broadening the lines and decreasing the sensitivity of the spectrum to the velocity fields. 

 Some inversion codes use the stray-light as a free parameter in the inversion procedure. In our opinion, stray light should be consistently applied to all the pixels: either by inferring it separately, before the inversion, or by simultaneously inverting the whole field-of-view \citep[similarly to, e.g.][]{2012A&A...548A...5V}. In this work, we do not employ any of these, specifically because we want to see how does the unknown stray light influence the inferred atmospheres.
 
\begin{figure}[htbp]
     \centering
     \includegraphics[width=0.5\textwidth]{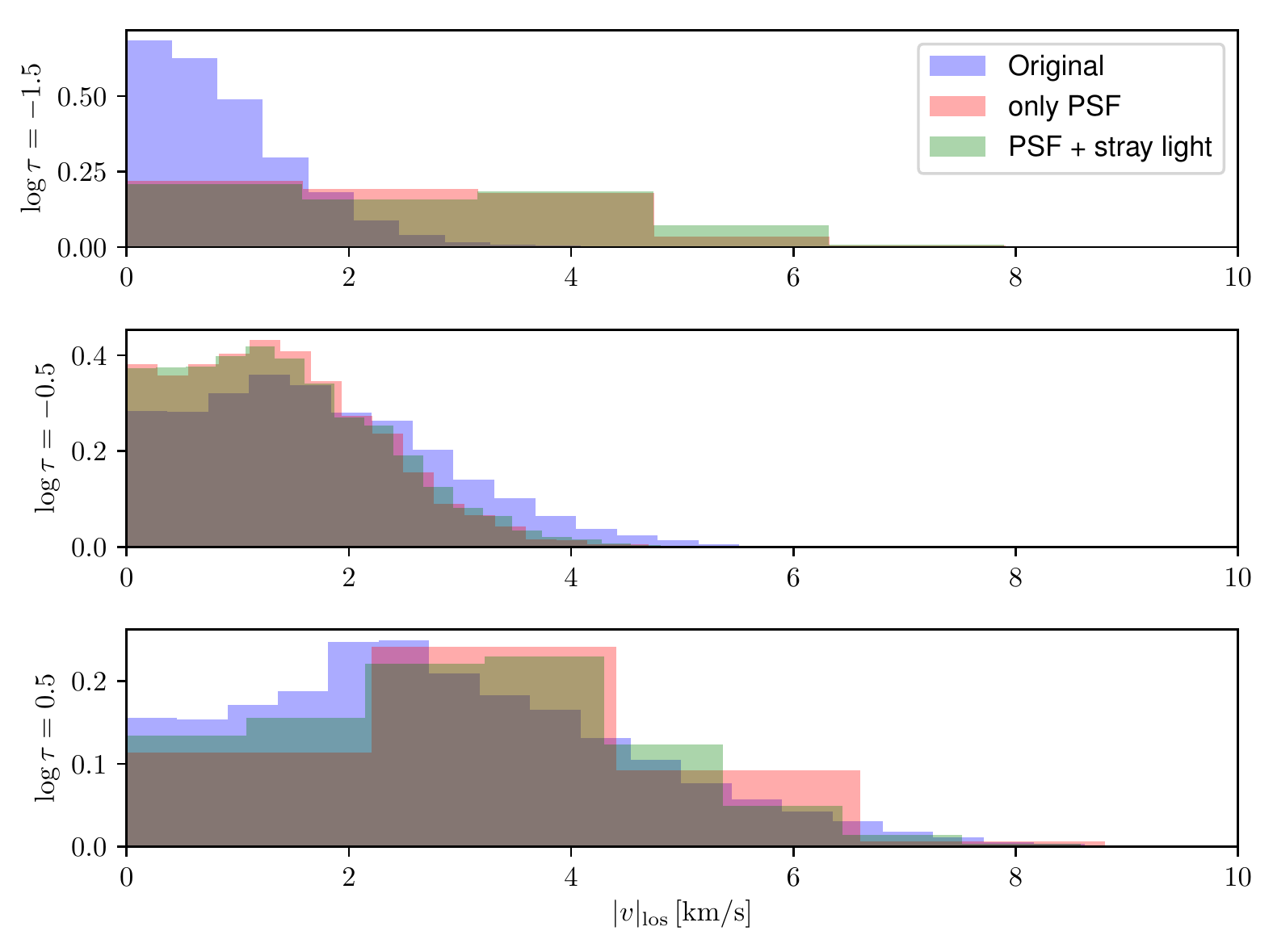}
     \caption{Histograms of the absolute values of line of sight velocities in the original atmosphere and as inferred from inversion of PSF-smeared data, without and with the additional stray light.} 
     \label{fig:histograms}
 \end{figure}

 Qualitatively, presence of stray light in these amounts does not fundamentally change the inferred velocities, and the inferred velocity maps (Fig.\,\ref{fig:v_comparison_2}) exhibit the same properties as discussed in \ref{ssection:psf}. As a way of quantitatively analyzing the data, we show the distributions of the absolute values of LOS velocities at these three atmospheric depths, for original MHD atmosphere, inversion of the spatially smeared data, and the inversion of spatially smeared data with the PSF. The distributions are given in Fig.~\ref{fig:histograms}. It seems like the presence of the stray light does not significantly change the inferred velocity distributions. The biggest difference is in the deepest layers which makes sense, because velocities there leave imprint in the far wings of spectral lines, where the intensity contrast is the lowest, and thus suffers the most from the presence of the stray light.

\section{Conclusions}
\label{sec:conclusions}

In this paper we studied the diagnostic potential of the five infrared iron lines around $15000\rm{\,\AA}$. The main premise being, because of the low continuum opacity around these wavelengths, we should be able to probe atmospheric parameters in deeper layers in comparison to the commonly used lines in the optical domain (e.g. Fe  6300$\,\AA$ line pair). In this particular study we focused on the LOS velocities, as it is of great interest to study the behaviour of convection at different atmospheric depths. 

To this end, we synthesized polarized spectra of the five considered spectral lines (Table~\ref{atomic_table}), using a state of the art MuRAM radiative magnetohydrodynamic simulation of the quiet solar atmosphere as the input. We then degraded the synthesized spectra by adding spectral PSF, photon noise, and the spatial, two-component PSF. By doing so, we have tried to mimic the telescope data which was subsequently inverted (i.e. tried to infer back the atmospheric parameters in the input MHD cube). We then directly compared the inferred and original velocities at three iso-optical depth surfaces ($\log\tau=-1.5,-0.5,0.5$). Our analysis indicates the following: 
\begin{itemize}
    \item Inversion of the 'original' (i.e. un-degraded) spectra, retrieves original velocities very well. 
    \item Spectrally smearing the data assuming modest spectral resolution of $10^5$ and adding a low amount of photon noise of $3\times10^-4$ in the units of continuum intensity, does not change results significantly.
    \item The agreement breaks at $\log\tau=1$, although the topology of the velocity field is roughly retrieved.
    \item Repeating the same experiment using $6300\,\AA$ lines of Iron shows that optical lines are much less sensitive to the deeper layers, though they have a non-zero response function at these layers. Namely, the intergranular velocities at deeper layers ($\log\tau=0.5$) are significantly overestimated. In the layers where $-1.5<\log\tau<0$, in the absence of spatial PSF, both spectral regions show excellent diagnostic potential.
    \item Inversion of spatially smeared, and binned data leads to the expected loss of spatial detail but it also seems that the velocities in the upper layers are severely misdiagnosed: granular velocities seem to be higher than the original ones, while the area covered by downflows seems to be smaller. 
    \item Finally, adding another, wide component to the PSF, that mimics the stray light in the amount of $30\%$, does not change the inferred velocities significantly. This is evident both from Figs.\,\ref{fig:v_comparison_2} and \ref{fig:histograms}.
\end{itemize}

We hope that this investigation emphasizes the strengths of this spectral region in spectropolarimetric inversions. Estimating velocity (and the magnetic field) deep in the photosphere, combined with the chromospheric diagnostics paints a complete picture of the solar atmosphere and allows us to characterize important phenomena (energy transport, reconnection) better. Also, there might be phenomena happening only in the deep layers which we will not be able to detect using other diagnostics. Finally, with the advent of next generation solar telescopes, \citep[e.g. DKIST and EST][]{2014SPIE.9147E..07E, 2013MmSAI..84..379C}, it will be of crucial importance to use spectral regions outside of visible, and we argue that this is one of the very useful ones for the photospheric diagnostics.

\begin{acknowledgements}
We thank T. Riethm\"{u}ller for kindly providing the MHD cube used in this paper.  This project has received funding from the European Research Council (ERC) under the European Union's Horizon 2020 research and innovation programme (grant agreement No. 695075). This research has made use of NASA's Astrophysics Data System.
\end{acknowledgements}

\appendix
\section{Examples of fits and inferred profiles}
\label{Appendix A}

An interesting aspect of the paper is the question of how well we can fit the synthetic spectra from the MHD model atmospheres that have complicated stratifications. On the other hand, the model we use is based on a fixed number of nodes, which means that we are, inevitably, simplifying the model atmospheres. Since the noise we are using in this numerical experiment is very low ($3\times10^{-4}$), it is not a big surprise that we fail to obtain optimal fits. A simple criterion for an `optimal fit' in this case could be $\chi^2_{\rm red}\approx 1$, where $\chi^2_{\rm red}$ is the reduced chi-squared).

\begin{figure*}[htbp]
     \centering
     \includegraphics[width=0.49\textwidth]{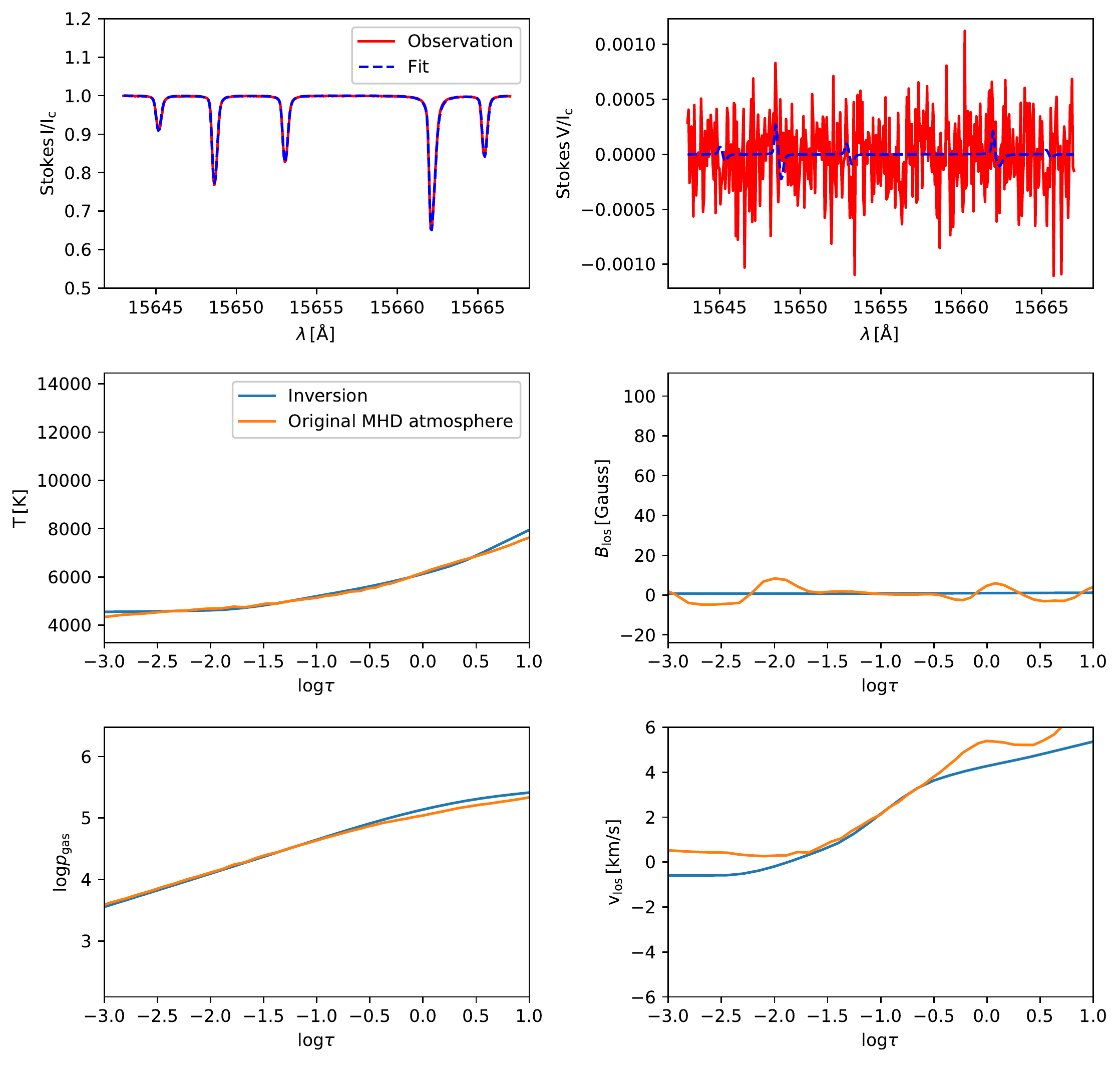}
     \includegraphics[width=0.49\textwidth]{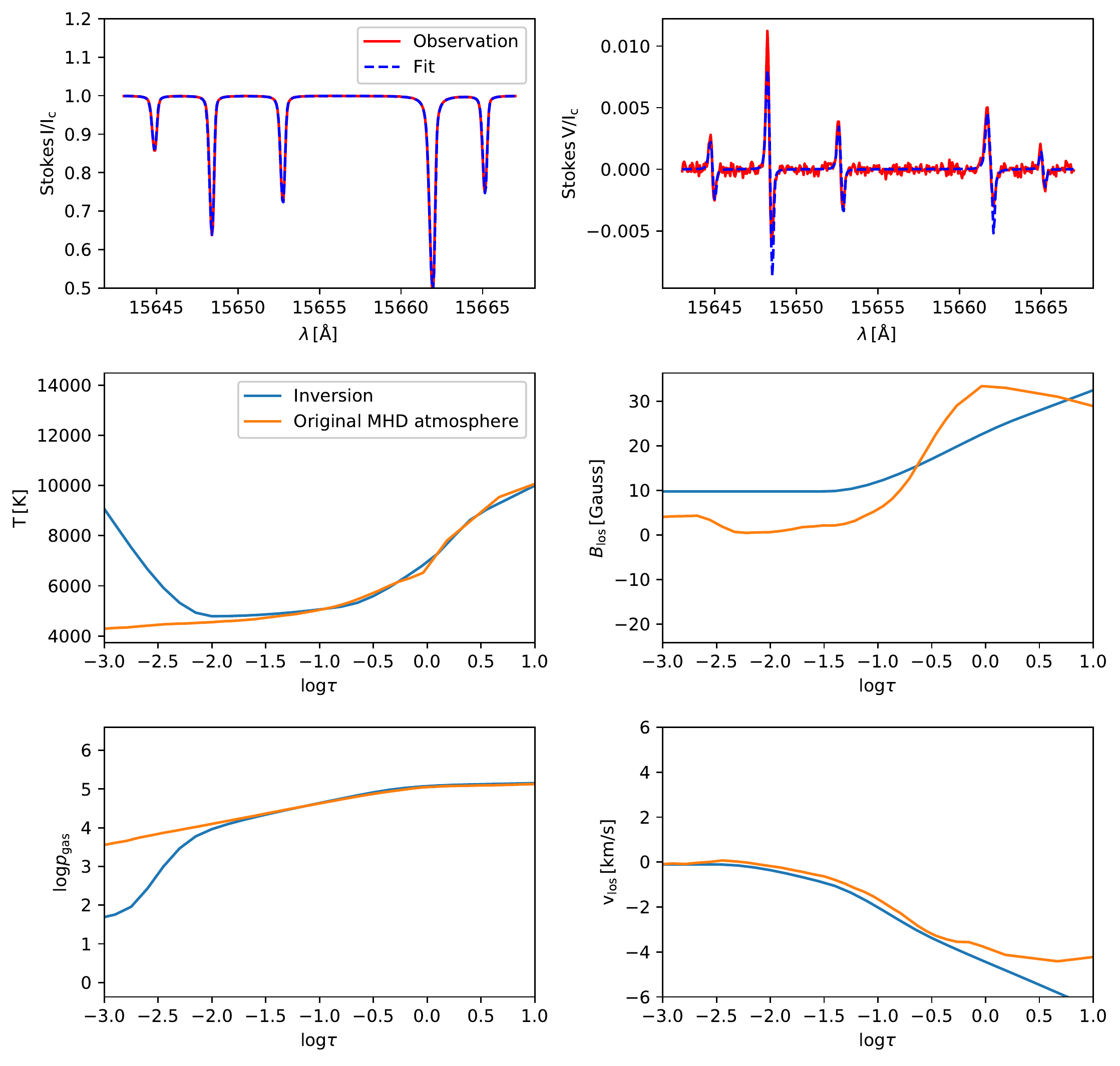}
     \caption{Upper four panels: Synthetic spectra and the best fits using node-based atmosphere model. Bottom eight panels: inferred and the original model atmospheres. Left: pixel with no polarization signal. Right: pixel with a weak polarization signal.} 
     \label{fig:App1.1}
 \end{figure*}

We show, in figure \ref{fig:App1.1} two representative fits and the infferred and the original atmospheres. Qualitatively speaking, Stokes $I$ fits are excellent, as continuum level, line depth and line shape are completely reproduced. Stokes $V$ fits are less good, but mostly because the magnitude of the Stokes $V$ is very small (the magnetic field is rather weak). It can be seen that temperature, gas pressure and the velocity are well reproduced, while the magnetic field stratification seems to be too complicated to be picked up with our simple, two-node model.

 \begin{figure}[htbp]
     \centering
     \includegraphics[width=0.5\textwidth]{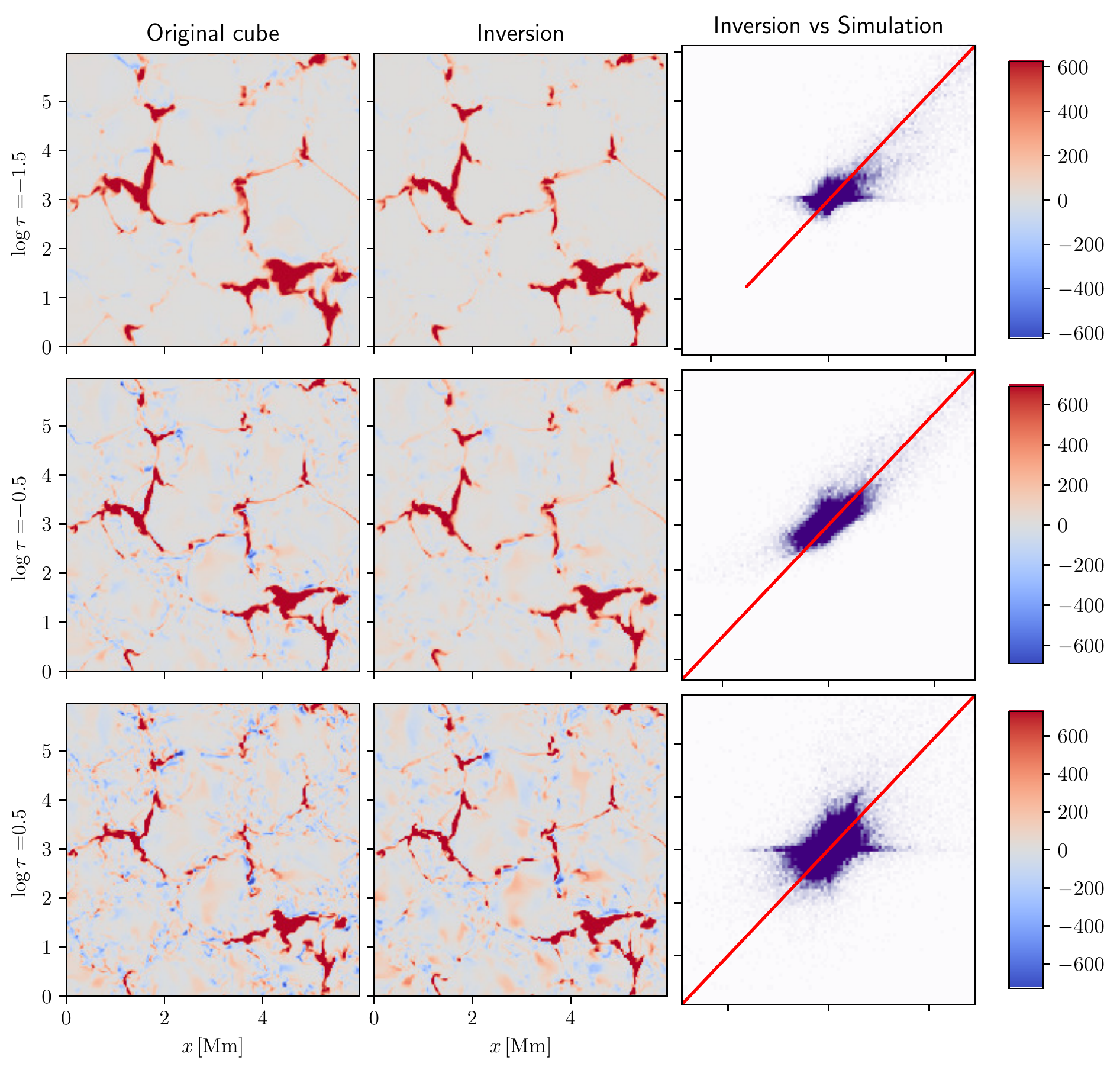}
     \caption{Comparison between line of sight magnetic fields in the MHD simulation and as returned by inversion code, from the spectrally degraded synthetic data, at three different depths in the atmosphere ($\log\tau = -1.5, -0.5, 0.5$). In all the figures, the magnetic field is in Gauss.}
     \label{fig:B_comparison_1}
 \end{figure}

However, the spatial distribution of the magnetic field seems to be well-recovered (see Fig.\,\ref{fig:B_comparison_1}). Qualitatively, the magnetic fields agree with the original ones, but because they are generally weak, there are many pixels where the discrepancy is big. We wanted to show these additional plots to demonstrate that our inversion scheme, although relatively simple, does yield good fits and reproduce well the original MHD cube. To recover very complicated depth stratifications, we might have to turn to more sophisticated methods, such as regularization methods as suggested by, for example, \citet{2018arXiv181008441D}. We note that there is a fundamental limit in height resolution, dictated by the radiative transfer processes that sets the limit on how fine vertical structure we can recover.

In this manuscript we chose to focus on convective (line of sight) velocities, but a similar study is in place regarding the magnetic fields, especially in the age of new solar telescopes that are supposed to provide us with a leap in spatial resolution and the polarimetric sensitivity.

\bibliography{ms}


\end{document}